# Particulate Generation on Surface of Iron Selenide Films by Air Exposure

Hidenori Hiramatsu[1,2,*], Kota Hanzawa[1], Toshio Kamiya[1,2], and Hideo Hosono[1,2]

[1] Laboratory for Materials and Structures, Institute of Innovative Research, Tokyo Institute of Technology, Mailbox R3-3, 4259 Nagatsuta-cho, Midori-ku, Yokohama 226-8503, Japan

[2] Materials Research Center for Element Strategy, Tokyo Institute of Technology, Mailbox SE-1, 4259 Nagatsuta-cho, Midori-ku, Yokohama 226-8503, Japan

*E-mail: h-hirama@mces.titech.ac.jp

Abstract

Nanometer-sized particular structures are generated on the surfaces of FeSe epitaxial films directly after exposure to air; this phenomenon was studied in the current work because these structures are an obstacle to field-induced superconductivity in electric double-layer transistors using FeSe channel layers. Chemical analyses using field-effect scanning Auger electron spectroscopy revealed no clear difference in the chemical composition between the particular structures and the other flat surface region. This observation limits the possible origins of the particulate formation to light elements in air such as O, C, H, and N.

1. Introduction

High critical temperature ($T_c$) iron-based superconductors have attracted substantial attention since reports of the superconductivity of 1111-type LaFePO [1] and fluorine-doped LaFeAsO [2]. These materials are interesting from the viewpoint of condensed matter physics and materials science [3, 4] while also possessing suitable properties for superconducting wire/tape/coated conductor applications [5–9], such as high field magnets and electric power cables. Therefore, many superconducting wires/tapes [10–14] and coated conductors [15–24] have been developed and demonstrated using iron-based superconductors. Their performance under high magnetic fields has rapidly improved and is approaching and/or overcoming that of commercially available conventional superconductors such as $Nb_3Sn$ and Nb-Ti.

One of the iron-based superconductors is 11-type FeSe [25], which has the simplest chemical composition among the materials. It is worth noting that $T_c$ of FeSe and related FeSe-based materials have been greatly enhanced using various techniques even though the $T_c$ of bulk FeSe is as low as ~8 K. For example, the application of an external high pressure [26] and chemical doping/intercalation of alkali metals and molecules [27–32] enhanced $T_c$ to approximately 40 K. The most remarkably enhanced $T_c$ of ~100 K was reported in monolayer-thick FeSe layers [33]. These results indicate that $T_c$ of FeSe is quite sensitive to change in the local structure and/or strain. The significant $T_c$ enhancements compared with that of the bulk are unique and reported only in FeSe and FeSe-related materials among the iron-based superconductors. Thus, FeSe is the most promising material for attaining high $T_c$ among iron-based layered arsenide/selenide materials.

Another type of $T_c$ enhancement of FeSe has been reported by employing electric double-layer transistor (EDLT) configurations with ionic liquid as a gate dielectric material [34–39]. In these recent reports, the EDLT configuration serves two roles: it allows electrostatic doping of high-density carriers to the EDLT channels [35–37], and electrochemical etching of the EDLT channels under relatively high gate bias (e.g., ≥ 5 V) and at high temperature (close to room temperature) is applied to control the channel thickness and electrical properties [34, 38, 39]. Both of these factors contribute to achieving a high $T_c$ of approximately 40 K, which is close to that for the above cases of high pressure and chemical doping/intercalation.

In our FeSe EDLT study in 2016 [36], we observed that our initial EDLTs did not work (i.e., no modulation of the channel resistance was observed by applying a gate bias) when the FeSe channels were exposed once to air even for a short time. Thus, we had to develop and employ an in situ sample-transfer system in ultrahigh vacuum or inert atmospheres [40] to fabricate EDLTs that exhibited field-induced superconductivity. This result indicated that the surface of the FeSe films is sensitive to air exposure. However, to our knowledge, degradation of the FeSe film surface and its detailed analysis have not yet been reported, even though the first report on superconductivity of FeSe was over 10 years ago in 2008 [25].

In this study, the effects of air exposure on the surface degradation of FeSe epitaxial films were examined. We were not able to detect a chemical difference between the degraded and not-degraded surfaces; however, this result implies the possible remaining origins of the degradation.

## 2. Experimental Details

In this work, ~10-nm-thick FeSe films were grown heteroepitaxially on (001)-oriented SrTiO$_3$ (STO) single crystals (size: 10 × 10 mm$^2$, 0.5-mm thick) using molecular beam epitaxy (MBE, EV-100/PLD-S made by Eiko Co.). Details of the film growth conditions and EDLT structure are provided in Refs [36,37]. The FeSe films were stored in an Ar-filled glove box (MDB-1BKHTMK type 2, Miwa Manufacturing Co., Ltd.), which was directly connected with the load lock/preparation chamber of the MBE system. The oxygen concentration and dew point of the glove box were << 1 ppm and approximately −100 °C, respectively.

An EDLT device was fabricated using a film patterned with a shadow mask and exposed to air for a few tens of minutes, and then, the temperature dependence of the sheet resistance ($R_s$) of the EDLT channel under a gate bias ($V_G$) was measured to determine the effect of the air exposure. Structural and chemical analyses (i.e., except the EDLT) were performed using 10 × 10-mm$^2$ samples.

$\theta$-Coupled $2\theta$ scan X-ray diffraction (XRD, SmartLab, Rigaku Co.) using CuK$\alpha$ radiation with a rotation Cu anode and power of 45 kV × 200 mA was performed using Bragg–Brentano geometry to determine the crystalline phases of the films. Note that the Bragg–Brentano geometry is unsuitable for evaluating epitaxial films because the X-ray scattering angle and line-width are large for the crystallite orientation distribution and diffraction peak widths of high-quality epitaxial films; therefore, the monochromatic parallel beam configuration is usually used. However, the use of the Bragg–Brentano geometry is an effective method to detect small amounts of impurity phases because we can use a blighter incident X-ray beam than that of the monochromatic parallel one.

XRD patterns of the as-grown films were obtained without air exposure using an O-ring sealed Ar-filled measurement holder to avoid surface degradation during the measurements. However, atomic force microscopy measurements (AFM, MultiMode 8 made by Bruker Nano Inc.), which were performed to observe the surface morphology, could not be conducted without air exposure. To transfer the films from the load lock/preparation chamber of the MBE apparatus to the AFM stage required ~200 s after air exposure of the films. However, only ~30 s was needed to transfer the film from the glove box to the AFM stage because we removed the sample from the substrate holder in the glove box (i.e., not in air) in this case. The surfaces of the films were exposed to air during AFM observation for 260 s per single scan.

The chemical composition of the surface was analyzed using field-emission scanning auger electron spectroscopy (FEAES, PHI 700, ULVAC PHI Inc.), in which four elements (C, O, Fe, and Se) were focused on under electron acceleration conditions of 10 kV and 10 nA to minimize damage by the electron beam. The observation points were selected using the scanning electron microscopy (SEM) mode of the PHI 700 instrument. The specification spatial resolution of the PHI 700 is 6 nm in SEM mode and 8 nm in AES mode. For the actual observation conditions, the lateral beam size and analysis depth (inelastic mean free path [IMFP] of emitted electrons) were estimated to be 20–30 and 3 nm, respectively.

The temperature dependences of the electrical resistivity ($\rho$) of the as-grown and air-exposed films were measured using the four-probe method using a physical property measurement system.

## 3. Results and Discussion

### 3.1 Generation of Nano-Sized Particular Structures at Film Surface by Air Exposure

Figures 1(a)–(c) present AFM images of the FeSe film, which was transferred from the load lock/preparation chamber to the AFM stage directly after film growth. We repeated the AFM scans three times in the same observation area. As observed in the cross-section (Fig. 1(d), the line indicated by the black and red arrows in Fig. 1(a)), two particular structures were observed in the first scan. One structure had a lateral size of ~20 nm and a height of ~3 nm at the position = 235 nm (indicated by the red vertical arrow); the other structure had a lateral size of ~20 nm and height of ~1 nm at the position = 87 nm (indicated by the black vertical arrow). Then, in the second scan image, the particular structure at position = 78 nm grew to a vertical size of ~4 nm (the black arrow), whereas the size of the particular structure at the position = 235 nm in the first scan image did not change. In the third scan, the sizes of both particular structures were almost the same; however, the observed position of one of them (that indicated by the black arrow) slightly changed for each scan. This result indicates that the small-sized particular structures grew to ~4 nm height within 260 s (= 1 scan time) and that adsorption of the particular structures to the surface would be very weak (i.e., the AFM probe would scratch and slightly manipulate them.). It should be noted that such fast air-degradation was not observed in another FeSe-related material, $TlFe_{1.6}Se_2$ [41]. Therefore, this phenomenon is unique to FeSe. We also examined other locations (the lines indicated by the green, blue, and pink arrows) during the second and third scans, as shown in Fig. 1(e). At this point, no significant change in the size or structure was observed. These results imply that the generation and growth of particular structures at the FeSe film surface are fast processes occurring within 260 s and that their growth

almost appears to be saturated at a height of ~4 nm at this time scale (within 980 s).

Next, we examined the effect of exposure to air for longer times of up to 1 h. The sample was stored in an Ar-filled glove box for 5 days, transferred to the AFM stage in air, and the AFM observation was started immediately. In the first scan image (Fig. 2(a)), almost the same size of the particular structures as that observed in Fig. 1(a) was observed. However, the density of the structures in Fig. 2(a) looks higher than that in Fig. 1(a), suggesting that the degradation rate was suppressed but that moderate degradation proceeded even in the Ar-filled glove box. After exposure to air for 30 min (Fig. 2(b)) and 60 min (Fig. 2(c)), the size of the particular structures became larger in both the lateral and height directions. The lateral sizes of the structures were ~20 nm for the 30-min sample and 30 nm for the 60-min sample. However, the lateral coverage of the particular structures almost remained the same. Compared with the result in Fig. 1(c,e) (after ~16 min), the size almost doubled after air exposure for 1 h. This result implies that the growth rate of the particular structures was low, whereas they gradually grew at the initial growth points with air exposure.

3.2 Investigation of Origin of Nano-Sized Particular Structures

To determine the origin of the particular structures at the FeSe film surface, we first performed XRD measurements of the air-exposed films (Fig. 3). However, no additional impurity phase was observed. We also confirmed that the *c*-axis lattice parameter did not change by comparing the peak positions of the FeSe 004 diffraction peak at $2\theta \approx 68.8°$ of the as-grown and air-exposed films (see the vertical dotted line in Fig. 3). However, the full width at half maximum (1.03°) of the FeSe 004 diffraction peak of the air-exposed film was wider than that (0.96°) of the as-grown film, suggesting that slight disorder of the layered structure and/or a slight decrease in the crystallite size occurred with air exposure. However, we did not detect any indication of this disorder or decrease in the crystallite size from the crystal structure or crystalline phase from the XRD measurements.

We then performed chemical composition point analysis of the FeSe film exposed to air for 150 min using FEAES. We observed many particular structures on the FeSe film surface in the SEM image (Fig. 4(a)), and then selected the two observation points indicated by the circles. One point was the position of a particular structure, and the other point was in the flat surface region without any particular structure. The lateral size of the particular structure was ~100 nm, which is larger than that of the sample exposed for 60 min (Fig. 2(c)), mainly because of the longer exposure time. The diameter of the incident electron beam and the IMFP of the emitted electrons under the observation condition were estimated to be 20–30 and 3 nm, respectively. These conditions guarantee that the electron beam is irradiated only within the particular structure and that the AES signal comes only from the particle structure. Figures 4(b) and (c) present the FEAES spectra of both regions. Although clear signals were observed for the main constituent elements (Fe and Se), we also observed significant contamination by carbon and oxygen. This result indicates that the entire top surface of the FeSe film exhibits similar air-sensitivity, regardless of the presence of the particular structures. Although the spectra were similar, we performed subtraction between (b) and (c). As indicated by the two vertical arrows, small difference peaks were observed for the C and O signals, whereas no difference was detected for Fe and Se. We also confirmed using SEM that the particular structures disappeared in the FEAES measurement chamber for longer observation times or higher acceleration voltages. We cannot propose the exact origin of the particular structures from the FEAES results; however, these results suggest that the generated structures were composed of and reacted with light elements in air such as O, C, H, and N.

The above speculation for the origin (i.e., light elements in air) is also supported by our findings regarding specimen preparation for transmission electron microscopy (TEM) observation. We attempted to fabricate TEM specimens using a film exposed to air for 3 h that was covered with a gold film as a protection layer using a standard thinning technique combined with cutting, bonding, ion milling, and dimpling. However, we could not find the particular structures in the TEM specimens (the usual lateral size in the thinned region of TEM specimens is a few $\mu m^2$), implying that the particular structures easily disappeared with mechanical contact (during the cutting and bonding processes) and/or ion damage (during milling and dimpling) and not only with irradiation of the electron beam.

3.3 EDLT Using Air-Exposed Channel Layer

Next, we investigated the effect of the surface degradation on the EDLT properties. The temperature dependence of the sheet resistance ($R_s$) was measured for an EDLT channel using an FeSe film air-exposed for a few tens of minutes (Fig. 5). This initial insulator-like $R_S$ behavior is similar to that of the EDLT without exposure to air reported in Refs. [36,37], although the $R_S$ behavior in the low-temperature region was different. However, even though a high $V_G$ of +4.0 V was applied, no phase transition was detected, and $R_s$ monotonously increased over the entire temperature range; clear modulation of the drain current and phase transition were not observed. These findings suggest that the particular structures generated at the surface act as an obstacle to field-induced phase transition and superconductivity in the FeSe EDLTs. For this reason, electrochemical etching [34, 38, 39], which should be able to remove the particular structures, is effective for achieving high-$T_c$ EDLTs if the FeSe channels must be exposed to air during the device fabrication process.

3.4 Long-Time Stability in Glove Box

Finally, we discuss the long-time stability of FeSe films. Figure 6 summarizes the results for the film stored in a glove box for ~2 years. Figures 6(a) and (b) present XRD patterns of the film. In this case, no additional impurity phase was detected, as shown in Fig. 6(a). However, compared with the as-grown film, the $c$-axis of the long-time-stored film expanded and approached the bulk value, as indicated by the vertical dotted line in (b). This result indicates that the film stored for a long time was almost fully relaxed similar to the bulk FeSe sample. Figs. 6 (c)–(e) show the surface morphology of the film stored for a long time. Although the film was stored in a glove box, much larger particular structures were observed than those in Figs. 1 and 2. This result suggests that a very small amount of oxygen or related air-constituting element can enhance the growth of the structures, although the glove box maintains a very-high-purity inert and dry atmosphere with an oxygen concentration << 1 ppm and a dew point of approximately −100 °C. Figure 6(f) shows the temperature dependence of the resistivity ($\rho - T$) curves. Air exposure for 60 min slightly increased $\rho$. However, a remarkable increase in $\rho$ was observed for the sample stored for a long time, especially in the low-temperature region (a hump structure is observed at ~30 K in the $\rho$–$T$ curve of the sample stored for 2 years; however, the origin is currently unclear). Before this experiment, we expected that superconductivity may be observed because the lattice parameter is fully relaxed, as observed in Fig. 6(a,b). However, we did not observe superconductivity. Therefore, surface contamination (i.e., the particular structures) dominated the $\rho$–$T$ behavior of the sample stored for 2 years.

4. Summary

We observed the generation of particular structures at the surface of FeSe epitaxial films directly after exposure to air. The origin of their generation remains unclear because no clear difference was detected in the chemical compositions of the particular structures and the flat bulk region. We propose that the origin is related to light elements in air such as O, C, H, and N based on the Auger electron spectroscopy results. An electric-field-induced phase transition was not observed for the EDLT using the air-exposed FeSe unlike for previously reported non-air-exposed FeSe EDLTs [36,37], and the resistance increased monotonously under the application of a positive gate bias. These results indicate that an in situ process without exposure to air [40] is necessary for electric-field-induced superconductivity in FeSe EDLTs if an electrochemically etching process [34,38,39] is not employed.

Funding information

This work was supported by the Ministry of Education, Culture, Sports, Science,

and Technology (MEXT) through the Element Strategy Initiative to Form Core Research Center. H. Hi. was also supported by the Japan Society for the Promotion of Science (JSPS) through Grant-in-Aid for Scientific Researches (A) and (B) (Grant Nos. 17H01318 and 18H01700), and Support for Tokyotech Advanced Research (STAR).

References

1. Kamihara, Y., Hiramatsu, H., Hirano, M., Kawamura, R., Yanagi, H., Kamiya, T., Hosono, H.: Iron-Based Layered Superconductor: LaOFeP. J. Am. Chem. Soc. **128**, 10012 – 10013 (2006).
https://doi.org/10.1021/ja063355c

2. Kamihara, Y., Watanabe, T., Hirano, M., Hosono, H.: Iron-Based Layered Superconductor La[$O_{1-x}F_x$]FeAs ($x$ = 0.05–0.12) with $T_c$ = 26 K. J. Am. Chem. Soc. **130**, 3296 – 3297 (2008).
https://doi.org/10.1021/ja800073m

3. For a review, Hosono, H., Kuroki, K.: Iron-based superconductors: Current status of materials and pairing mechanism. Physica C: Superconductivity and its Applications **514**, 399 – 422 (2015).
https://doi.org/10.1016/j.physc.2015.02.020

4. For a review, Hosono, H., Tanabe, K., Takayama-Muromachi, E., Kageyama, H., Yamanaka, S., Kumakura, H., Nohara, M., Hiramatsu, H., Fujitsu, S.: Exploration of new superconductors and functional materials, and fabrication of superconducting tapes and wires of iron pnictides. Sci. Technol. Adv. Mater. **16**, 033503 (2015).
https://doi.org/10.1088/1468-6996/16/3/033503

5. For a review, Putti, M., Pallecchi, I., Bellingeri, E., Cimberle, M. R., Tropeano, M., Ferdeghini, C., Palenzona, A., Tarantini, C., Yamamoto, A., Jiang, J., Jaroszynski, J., Kametani, F., Abraimov, D., Polyanskii, A., Weiss, J. D., Hellstrom, E. E., Gurevich, A., Larbalestier, D. C., Jin, R., Sales, B. C., Sefat, A. S., McGuire, M. A., Mandrus, D., Cheng, P., Jia, Y., Wen, H. H., Lee, S., Eom, C. B.: New Fe-based superconductors: properties relevant for applications. Supercond. Sci. Technol. **23**, 034003 (2010).
https://doi.org/10.1088/0953-2048/23/3/034003

6. For a review, Ma, Y.: Progress in wire fabrication of iron-based superconductors. Supercond. Sci. Technol. **25**, 113001 (2012).
https://doi.org/10.1088/0953-2048/25/11/113001

7. For a review, Shimoyama, J.: Potentials of iron-based superconductors for practical future materials. Supercond. Sci. Technol. **27**, 044002 (2014).
https://doi.org/10.1088/0953-2048/27/4/044002

8. For a review, Hosono, H., Yamamoto, A., Hiramatsu, H., Ma, Y.: Recent advances in iron-based superconductors toward applications. Mater. Today **21**, 278 – 302 (2018).
https://doi.org/10.1016/j.mattod.2017.09.006

9. For a review, Iida, K., Hänisch, J., Tarantini, C.: Fe-based superconducting thin films on metallic substrates: Growth, characteristics, and relevant properties. Appl. Phys. Rev. **5**, 031304 (2018).
https://doi.org/10.1063/1.5032258

10. Weiss, J. D., Tarantini, C., Jiang, J., Kametani, F., Polyanskii, A. A., Larbalestier, D. C., Hellstrom, E. E.: High intergrain critical current density in fine-grain $(Ba_{0.6}K_{0.4})Fe_2As_2$ wires and bulks. Nat. Mater. **11**, 682 – 685 (2012).
https://doi.org/10.1038/NMAT3333

11. Togano, K., Matsumoto, A., Kumakura, H.: Fabrication and transport properties of ex situ powder-in-tube (PIT) processed $(Ba,K)Fe_2As_2$ superconducting wires. Solid State Commun. **152**, 740–746 (2012).
https://doi.org/10.1016/j.ssc.2011.12.014

12. Gao, Z., Togano, K., Matsumoto, A., Kumakura, H.: Achievement of practical level critical current densities in $Ba_{1-x}K_xFe_2As_2$/Ag tapes by conventional cold mechanical deformation. Sci. Rep. **4**, 4065 (2014).
https://doi.org/10.1038/srep04065

13. Weiss, J. D., Yamamoto, A., Polyanskii, A. A., Richardson, R. B., Larbalestier, D. C., Hellstrom, E. E.: Demonstration of an iron-pnictide bulk superconducting magnet capable of trapping over 1 T. Supercond. Sci. Technol. **28**, 112001 (2015).
https://doi.org/10.1088/0953-2048/28/11/112001

14. Zhang, X., Oguro, H., Yao, C., Dong, C., Xu, Z., Wang, D., Awaji, S., Watanabe, K., Ma, Y.: Superconducting Properties of 100-m Class $Sr_{0.6}K_{0.4}Fe_2As_2$ Tape and Pancake Coils. IEEE Trans. Appl. Supercond. **27**, 7300705 (2017).
https://doi.org/10.1109/TASC.2017.2650408

15. Iida, K., Hänisch, J., Trommler, S., Matias, V., Haindl, S., Kurth, F., Lucas, del Pozo, I., Hühne, R., Kidszun, M., Engelmann, J., Schultz, L., Holzapfel, B.: Epitaxial Growth of Superconducting Ba(Fe$_{1-x}$Co$_x$)$_2$As$_2$ Thin Films on Technical Ion Beam Assisted Deposition MgO Substrates. Appl. Phys. Express **4**, 013103 (2011).
    https://doi.org/10.1143/APEX.4.013103
16. Katase, T., Hiramatsu, H., Matias, V., Sheehan, C., Ishimaru, Y., Kamiya, T., Tanabe, K., Hosono, H.: Biaxially textured cobalt-doped BaFe$_2$As$_2$ films with high critical current density over 1 MA/cm$^2$ on MgO-buffered metal-tape flexible substrates. Appl. Phys. Lett. **98**, 242510 (2011).
    https://doi.org/10.1063/1.3599844
17. Si, W., Zhou, J., Jie, Q., Dimitrov, I., Solovyov, V., Johnson, P. D., Jaroszynski, J., Matias, V., Sheehan, C., Li, Q.: Iron-chalcogenide FeSe$_{0.5}$Te$_{0.5}$ coated superconducting tapes for high field applications. Appl. Phys. Lett. **98**, 262509 (2011).
    https://doi.org/10.1063/1.3606557
18. Trommler, S., Hänisch, J., Matias, V., Hühne, R., Reich, E., Iida, K., Haindl, S., Schultz, L., Holzapfel, B.: Architecture, microstructure and $J_c$ anisotropy of highly oriented biaxially textured Co-doped BaFe$_2$As$_2$ on Fe/IBAD-MgO-buffered metal tapes. Supercond. Sci. Technol. **25**, 084019 (2012).
    https://doi.org/10.1088/0953-2048/25/8/084019
19. Si, W., Han, S. J., Shi, X., Ehrlich, S. N., Jaroszynski, J., Goyal, A., Li, Q.: High current superconductivity in FeSe$_{0.5}$Te$_{0.5}$-coated conductors at 30 tesla. Nat. Commun. **4**, 1347 (2013).
    https://doi.org/10.1038/ncomms2337
20. Iida, K., Kurth, F., Chihara, M., Sumiya, N., Grinenko, V., Ichinose, A., Tsukada, I., Hänisch, J., Matias, V., Hatano, T., Holzapfel, B., Ikuta, H.: Highly textured oxypnictide superconducting thin films on metal substrates. Appl. Phys. Lett. **105**, 172602 (2014).
    https://doi.org/10.1063/1.4900931
21. Sato, H., Hiramatsu, H., Kamiya, T., Hosono, H.: Enhanced critical-current in P-doped BaFe$_2$As$_2$ thin films on metal substrates arising from poorly aligned grain boundaries. Sci. Rep. **6**, 36828 (2016).
    https://doi.org/10.1038/srep36828
22. Iida, K., Sato, H., Tarantini, C., Hänisch, J., Jaroszynski, J., Hiramatsu, H.,

Holzapfel, B., Hosono, H.: High-field transport properties of a P-doped BaFe$_2$As$_2$ film on technical substrate. Sci. Rep. **7**, 39951 (2017). https://doi.org/10.1038/srep39951

23. Xu, Z., Yuan, P., Ma, Y., Cai, C.: High-performance FeSe$_{0.5}$Te$_{0.5}$ thin films fabricated on less-well-textured flexible coated conductor templates. Supercond. Sci. Technol. **30**, 035003 (2017).
    https://doi.org/10.1088/1361-6668/30/3/035003

24. Xu, Z., Yuan, P., Fan, F., Chen, Y., Ma, Y.: Transport properties and pinning analysis for Co-doped BaFe$_2$As$_2$ thin films on metal tapes. Supercond. Sci. Technol. **31**, 055001 (2018).
    https://doi.org/10.1088/1361-6668/aab261

25. Hsu, F.-C., Luo, J.-Y., Yeh, K.-W., Chen, T.-K., Huang, T.-W., Wu, P.M., Lee, Y.-C., Huang, Y.-L., Chu, Y.-Y., Yan, D.-C., Wu, M.-K.: Superconductivity in the PbO-type structure α-FeSe. Proc. Natl. Acad. Sci. USA **105**, 14262 – 14264 (2008).
    https://doi.org/10.1073/pnas.0807325105

26. Medvedev, S., McQueen, T. M., Troyan, I. A., Palasyuk, T., Eremets, M. I., Cava, R. J., Naghavi, S., Casper, F., Ksenofontov, V., Wortmann, G., Felser, C.: Electronic and magnetic phase diagram of *β*-Fe$_{1.01}$Se with superconductivity at 36.7 K under pressure. Nat. Mater. **8**, 630–633 (2009).
    https://doi.org/10.1038/nmat2491

27. Guo, J., Jin, S., Wang, G., Wang, S., Zhu, K., Zhou, T., He, M., Chen, X.: Superconductivity in the iron selenide K$_x$Fe$_2$Se$_2$ (0≤$x$≤1.0). Phys. Rev. B **82**, 180520(R) (2010).
    https://doi.org/10.1103/PhysRevB.82.180520

28. Ying, T. P., Chen, X. L., Wang, G., Jin, S. F., Zhou, T. T., Lai, X. F., Zhang, H., Wang, W. Y.: Observation of superconductivity at 30 – 46 K in A$_x$Fe$_2$Se$_2$ (A = Li, Na, Ba, Sr, Ca, Yb, and Eu). Sci. Rep. **2**, 426 (2012).
    https://doi.org/10.1038/srep00426

29. Burrard-Lucas, M., Free, D.G., Sedlmaier, S.J., Wright, J.D., Cassidy, S.J., Hara, Y., Corkett, A.J., Lancaster, T., Baker, P.J., Blundell, S.J., Clarke, S.J.: Enhancement of the superconducting transition temperature of FeSe by intercalation of a molecular spacer layer. Nat. Mater. **12**, 15 – 19 (2013).
    https://doi.org/10.1038/nmat3464

30. Guo, J., Lei, H., Hayashi, F., Hosono, H.: Superconductivity and phase instability of NH$_3$-free Na-intercalated FeSe$_{1-z}$S$_z$. Nat. Commun. **5**, 4756
15

(2014).

https://doi.org/10.1038/ncomms5756

31. Sedlmaier, S. J., Cassidy, S. J., Morris, R. G., Drakopoulos, M., Reinhard, C., Moorhouse, S. J., O'Hare, D., Manuel, P., Khalyavin, D., Clarke, S. J.: Ammonia-Rich High-Temperature Superconducting Intercalates of Iron Selenide Revealed through Time-Resolved *in Situ* X-ray and Neutron Diffraction. J. Am. Chem. Soc. **136**, 630–633 (2014).

    https://doi.org/10.1021/ja411624q

32. Lu, X. F., Wang, N. Z., Wu, H., Wu, Y. P., Zhao, D., Zeng, X. Z., Luo, X. G., Wu, T., Bao, W., Zhang, G. H., Huang, F. Q., Huang, Q. Z., Chen, X. H.: Coexistence of superconductivity and antiferromagnetism in $(Li_{0.8}Fe_{0.2})OHFeSe$. Nat. Mater. **14**, 325–329 (2015).

    https://doi.org/10.1038/nmat4155

33. Ge, J.-F., Liu, Z.-L., Liu, C., Gao, C.-L., Qian, D., Xue, Q.-K., Liu, Y., Jia, J.-F.: Superconductivity above 100 K in single-layer FeSe films on doped $SrTiO_3$. Nat. Mater. **14**, 285–289 (2015).

    https://doi.org/10.1038/nmat4153

34. Shiogai, J., Ito, Y., Mitsuhashi, T., Nojima, T., Tsukazaki, A.: Electric-field-induced superconductivity in electrochemically etched ultrathin FeSe films on $SrTiO_3$ and MgO. Nat. Phys. **12**, 42 – 46 (2016).

    https://doi.org/10.1038/NPHYS3530

35. Lei, B., Cui, J.H., Xiang, Z.J., Shang, C., Wang, N.Z., Ye, G.J., Luo, X.G., Wu, T., Sun, Z., Chen, X.H.: Evolution of High-Temperature Superconductivity from a Low-$T_c$ Phase Tuned by Carrier Concentration in FeSe Thin Flakes. Phys. Rev. Lett. **116**, 077002 (2016).

    https://doi.org/10.1103/PhysRevLett.116.077002

36. Hanzawa, K., Sato, H., Hiramatsu, H., Kamiya, T., Hosono, H.: Electric field-induced superconducting transition of insulating FeSe thin film at 35 K. Proc. Natl. Acad. Sci. USA **113**, 3986 – 3990 (2016).

    https://doi.org/10.1073/pnas.1520810113

37. Hanzawa, K., Sato, H., Hiramatsu, H., Kamiya, T., Hosono, H.: Key Factors for Insulator–Superconductor Transition in FeSe Thin Films by Electric Field. IEEE Trans. Appl. Supercond. **27**, 7500405 (2017).

    https://doi.org/10.1109/TASC.2016.2639738

38. Shiogai, J., Miyakawa, T., Ito, Y., Nojima, T., Tsukazaki, A.: Unified trend of superconducting transition temperature versus Hall coefficient for ultrathin

FeSe films prepared on different oxide substrates. Phys. Rev. B **95**, 115101 (2017).
https://doi.org/10.1103/PhysRevB.95.115101

39. Kouno, S., Sato, Y., Katayama, Y., Ichinose, A., Asami, D., Nabeshima, F., Imai, Y., Maeda, A., Ueno, K.: Superconductivity at 38 K at an electrochemical interface between an ionic liquid and Fe($Se_{0.8}Te_{0.2}$) on various substrates. Sci. Rep. **8**, 14731 (2018).
https://doi.org/10.1038/s41598-018-33121-7

40. Hiramatsu, H., Hosono, H.: Thin-film Growth and Device Fabrication of Iron-based Superconductors. TEION KOGAKU (J. Cryo. Super. Soc. Jpn) **52**, 433 – 442 (2017). (in Japanese)
https://doi.org/10.2221/jcsj.52.433

41. Katase, T., Hiramatsu, H., Kamiya, T., Hosono, H.: Electric double-layer transistor using layered iron selenide Mott insulator $TlFe_{1.6}Se_2$. Proc. Natl. Acad. Sci. USA **111**, 3979 – 3983 (2014).
https://doi.org/10.1073/pnas.1318045111

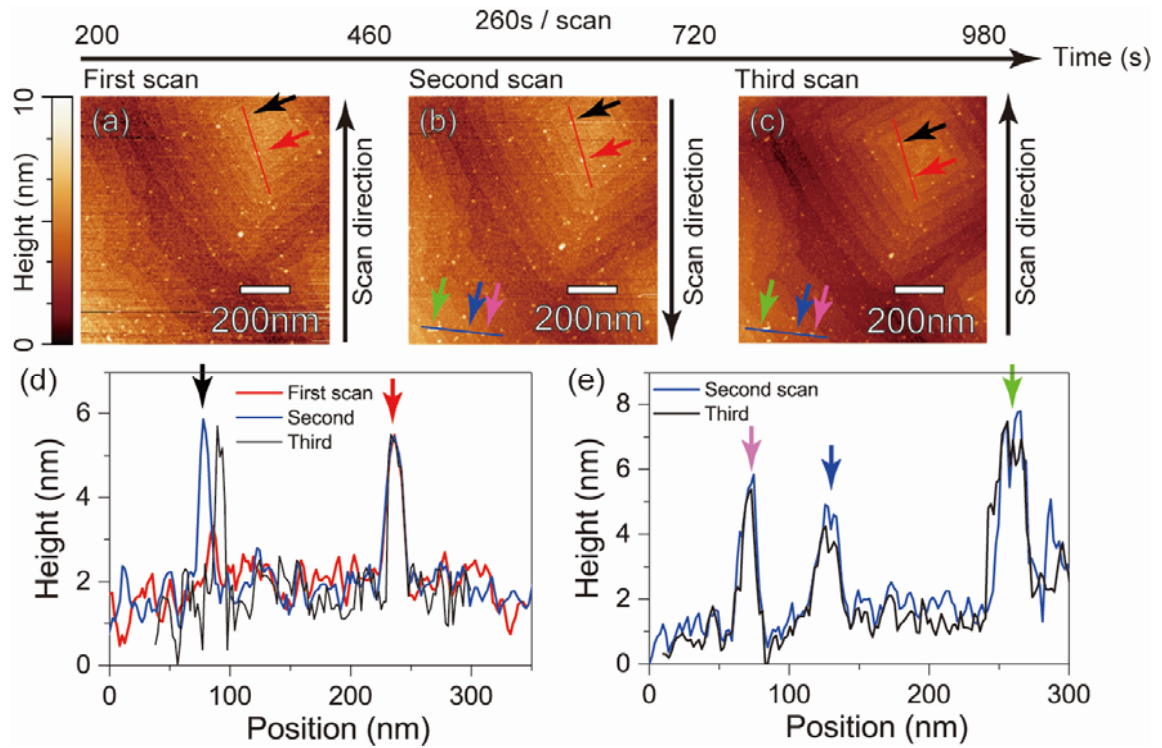

**Fig. 1** Surface morphology of FeSe film measured in air. (a)–(c) AFM images. The color bar indicates the height scale in the images. (a), (b), and (c) are results of a continuous repeating observation of the first, second, and third scans, respectively. The top horizontal arrow indicates the time after air exposure of the film, and the vertical arrows on the right of (a)–(c) are the scan direction of the AFM observation. The observation start time corresponds to 200 s because transferring the film from the load lock/preparation chamber of the MBE apparatus to the AFM stage required 200 s. (d) and (e) Cross-section profiles at the lines indicated by (d) vertical lines in (a)–(c) and (e) horizontal lines in (b)–(c).

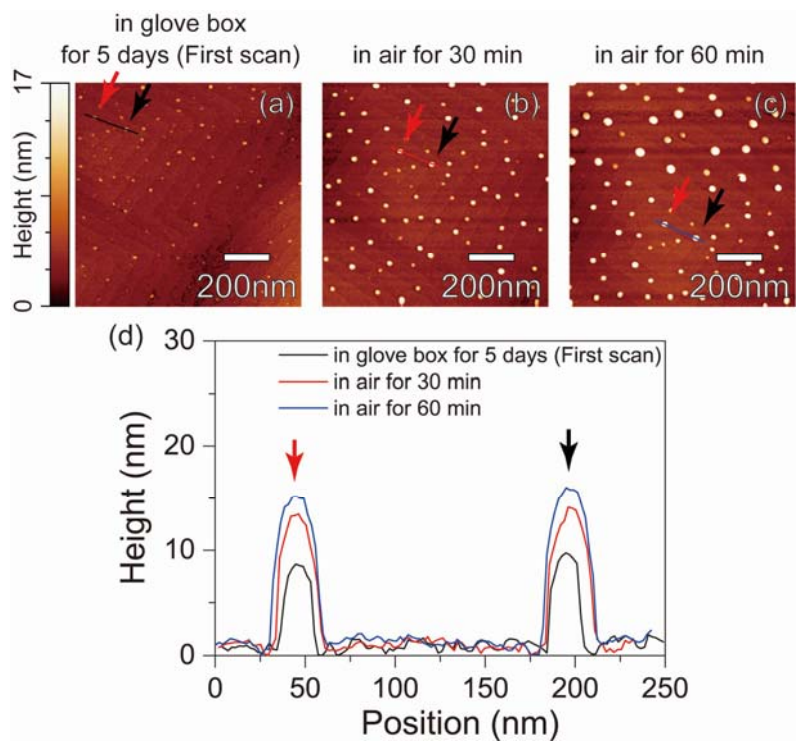

**Fig. 2** Change in surface morphology of air-exposed FeSe film for long times up to 1 h. (a)–(c) AFM images. The color bar indicates the height scale in the images. (a) First scan measured after storage for 5 days in Ar-filled glove box. (b) and (c) Images of sample (a) exposed to air for (b) 30 and (c) 60 min. (d) Cross-section profiles at lines in (a)–(c).

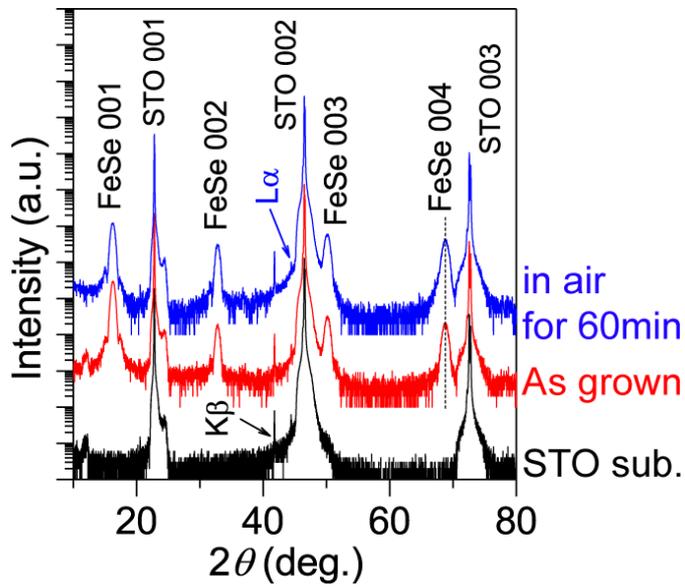

**Fig. 3** Out-of-plane XRD patterns of FeSe films. (Middle) As-grown film. These data were taken without air exposure using an Ar-filled measurement holder to avoid surface degradation during the measurements. (Top) Film with air exposure for 60 min. (Bottom) Pattern of bare STO single-crystal substrate for comparison. Weak diffraction peaks originating from Cu Kβ, W Lα, and some multiple reflections are observed from the single-crystal STO substrate.

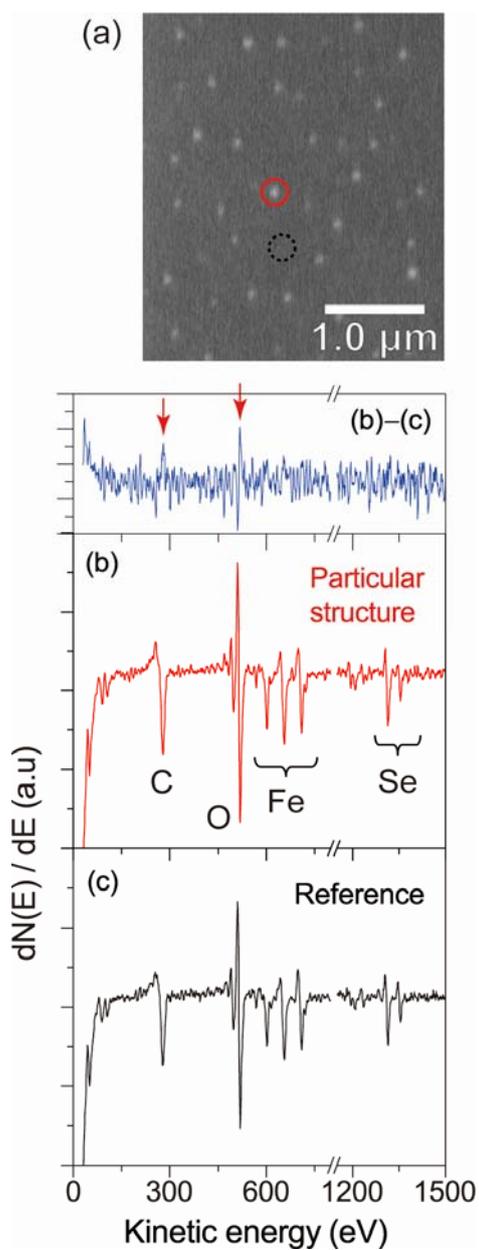

**Fig. 4** Chemical composition point analysis of FeSe film surface exposed to air for 150 min using FEAES. (a) SEM image of FeSe film to determine observation points. The red and the dotted black circles are the observation points corresponding to [(b): a particular structure position] and [(c): a flat surface position for reference]. The top figure over (b) is the subtraction spectrum between (b) and (c).

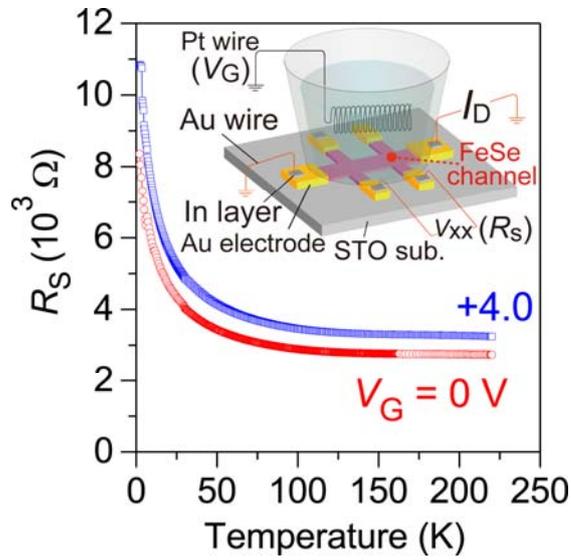

**Fig. 5** Temperature dependence of sheet resistance ($R_s$) of an EDLT device using an FeSe channel layer exposed to air for a few tens of minutes under gate bias voltages ($V_G$) of 0 and +4 V. The inset presents a schematic illustration of the EDLT.

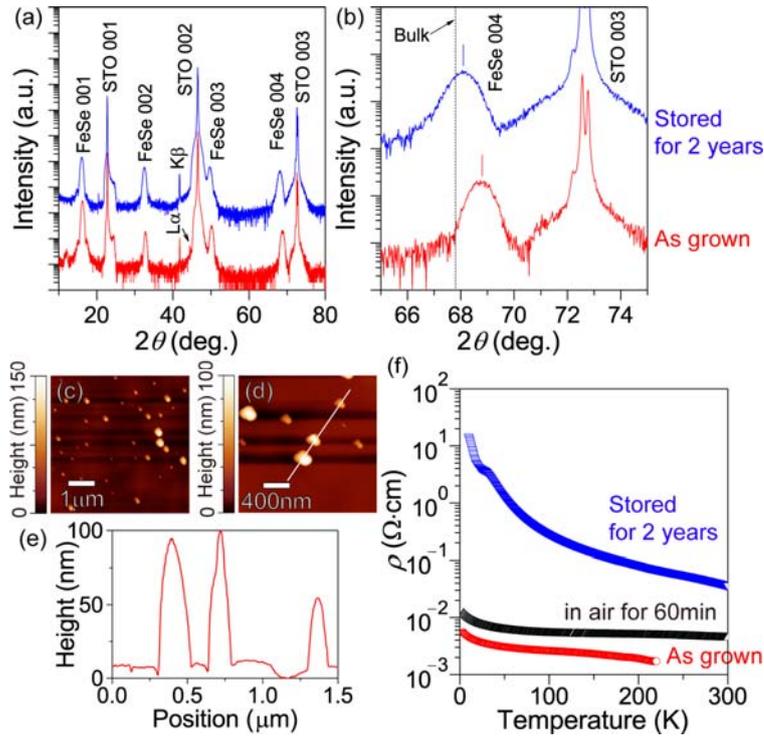

**Fig. 6** Structural and electrical properties of FeSe films stored in a glove box for 2 years. (a, b) Out-of-plane XRD pattern. (b) Enlarged pattern in the high $2\theta$ region. (c)–(e) Surface morphology. (c, d) AFM images of the film surface. The color bars show the height scale for each image. (d) Enlarged image of (c). (e) Cross-section profile of the line in (d). (f) Temperature dependence of resistivity ($\rho$) of the film. Data for the as-grown film and film exposed to air for 60 min are shown for comparison.